\documentstyle[aps,epsf,multicol]{revtex}
\begin{document}
\draft
\widetext
\title{How to observe distinct universal conductances in tunneling to
quantum Hall states: having the right contacts}
\author{Claudio de C. Chamon and Eduardo Fradkin}
\address{Department of Physics, University of Illinois at Urbana-Champaign,
Urbana, IL 61801-3080}
\maketitle
\begin{abstract}
We show that different {\it universal} values can be obtained for the 
two-terminal conductance of a fractional quantum Hall state.
At large voltages, or strong coupling, the conductance of a point-like
tunneling junction between an electron gas reservoir and a 
Laughlin FQH state at filling fraction $\nu$
{\it saturates} to a universal value $G={\frac{2\nu}{\nu+1}}
{\frac{e^2}{h}}$.
We use this result to show that devices with different types of contacts
between the reservoir and the FQH state lead to distinct universal 
values of saturation conductance which are rational multiples of $e^2/h$.
The particular fraction $e^2/h$ is obtained for the
case of electron tunneling in and out of a FQH liquid through two point 
contacts.
We demonstrate that the problem of tunneling between an electron gas 
and a fractional quantum Hall state through an impurity 
is exactly equivalent to the problem
of tunneling between a chiral Fermi liquid and a chiral Luttinger
liquid. We investigate in detail
the case of tunneling to a $\nu=1/3$ FQH state which we show to be
equivalent to the problem of tunneling {\it between} two $g=1/2$ chiral
Luttinger liquids. This system provides an experimental realization of 
this important exactly solvable case.
We use the results of the single impurity problem to consider
the case of many tunneling centers coupled independently to an electron
reservoir, which is relevant to recent experiments by A.~Chang 
{\it et.~al.} We derive an explicit universal expression for the 
voltage and temperature dependent conductance which 
exhibits a crossover reminiscent of a Kondo effect. This universal curve 
fits the experimental data over the full range of probed voltages.
\end{abstract}
\pacs{PACS: 73.23.-b, 71.10.Pm, 73.40.Hm, 73.40.Gk}


\begin{multicols}{2}
\narrowtext

\section{Introduction}
\label{sec:intro}

Since its discovery, the quantum Hall effect (QHE) has provided the
experimental setting for new ideas in condensed matter physics, as
well the inspiration for new ones. For example, the QHE establishes a
clean experimental realization of strongly correlated one-dimensional
Luttinger liquids. It was noted first in X.~G.~Wen's seminal work on
the theory of edge states of fractional quantum Hall (FQH) liquids
that the gapless edge modes were a realization of a (chiral) Luttinger
liquid \cite{XGWcll}. One of Wen's first proposals for testing these
ideas was to do experiments that would probe tunneling between the
edges of QH states\cite{XGWtun}. He predicted that in experiments in
QH {\it junctions}, the tunneling current should exhibit a power law
dependence on the applied voltage with an exponent determined by the
topological order of the bulk QH liquid. For a QH state with a single
edge the exponent is determined solely by the filling fraction $\nu$.
Tunneling in QH states is conceptually related to the problem of
scattering of quasiparticles in quantum wires (QW) which are
non-chiral Luttinger liquids.  C.~Kane and M.~P.~A.~Fisher have given
a physical picture of tunneling in Luttinger liquids in the form of a
renormalization group theory for both chiral and non-chiral liquids.
In particular, they predicted the nonlinear $I-V$ characteristics and
the universal shape of the conductance peaks \cite{Kane&Fisher1}.

Experimentally, the power law scaling of the tunneling current on
voltage and temperature in Luttinger liquids has been observed by two
groups. Milliken, Umbach and Webb \cite{Webb} used a gated quantum
point-contact to bring the edges of a FQH state closer so as to
observe the tunneling current. This is an experimental realization of
tunneling {\it between} the edge states of QH liquids.  More recently,
Chang, Pfeiffer and West \cite{Chang} measured electron tunneling from
a bulk doped-GaAs electron gas {\it into} the abrupt edge of a FQH
state. The latter experiment is particularly interesting for the
following reasons. First, it is able to observe the power law behavior
over more than one decade in $V$ and $T$.
Secondly, the device used in the experiment is an effective tool to
address experimentally the study of how a bulk electron gas serves as
reservoir when in contact to sharply defined FQH liquids.

QH tunneling junctions are an ideal setting to study the
physical properties of bulk QH liquids and, in particular, the nature
of their quasiparticles. These properties determine the power law
behavior of the nonlinear $I-V$ curves. This behavior of the
$I-V$ characteristics is just one of the many aspects of this rich
problem.

From the theoretical point of view, the problem of tunneling in chiral
Luttinger liquids is described by rich quantum field theories with a
non-trivial spectrum, exhibiting a weak-to-strong coupling duality
symmetry \cite{Duality}.  For special values of the Luttinger liquid
parameter $g$, the model is exactly solvable via the thermodynamic
Bethe ansatz \cite{Fendley1}. The conductance is known for all values
of $g$ in terms of dual series expansions for the values $g$ and $1/g$
\cite{Fendley1,Weiss}. The noise spectrum contains structures related
to the fractional charge and statistics of the tunneling particles
\cite{Kane&Fisher2,CFW1,Fendley2}, and it is believed to be a
non-analytic function of $g$ for finite frequencies
\cite{CFW2}. Hence, tunneling experiments to QH edges provide a unique
window to study the rich and deep physics of chiral Luttinger liquids
beyond the determination of their asymptotic scaling behavior.

Experimental devices similar to those of A.~Chang {\it et. al.\/} can
be used to address such conceptual issues.  Moreover, these devices
also provide a means for studying mesoscopic effects arising from
different ways of coupling the reservoirs (the bulk electron gas) to
the FQH states\cite{IQHcontacts}.

We have learned in recent years that the transport properties of
strongly correlated states depend delicately on the properties of the
contacts or reservoirs. This is the case for 1D quantum wires, which
display quantized conductances in agreement with non-interacting
electrons \cite{Tarucha,Maslov&Stone}, despite the renormalization
which should be expected from the electron-electron interactions in
infinite systems \cite{Apel&Rice}. It may seem {\it a priori} that
there should be a clear distinction between non-chiral Luttinger liquids
(quantum wires) and chiral ones (edge states), for the quantized Hall
conductance is not altered by the reservoirs \cite{Alekseev}. However,
we will see below that this is a rather subtle issue.

The physics of tunneling of {\it electrons} from a reservoir to the
edges of a QH liquid is determined by the physical properties of the
QH {\it quasiparticles}. The natural question that arises is whether
it is possible to use QH junctions to access the properties of the
quasiparticles of fractional quantum Hall states, in particular their
fractional charge.

The question of how to observe a fractionally charged quasiparticle of
a strongly correlated system is a rather old problem. It has been
considered in detail in the context of the soliton states in
quasi-one-dimensional conductors such as
polyacetylene\cite{Schrieffer}.  It has been proposed that noise
experiments in both polyacetylene\cite{Kivelson} and in QH
junctions\cite{Kane&Fisher2,CFW1} could be used to measure the
fractional charge. However, experimental attempts to measure noise in
both physical systems have encountered a number of significant
technical and conceptual difficulties.  Observing a
fractionally charged excitation from a bulk FQH state by tunneling
from a known state (the reservoir) depends on the nature of both the
reservoir and the contacts\cite{Landauer}.  In the context of the FQHE, the
proposed noise experiments rely on the theoretical picture in which
edge states of FQHE liquids are {\it assumed} to be in equilibrium
with a hypothetical reservoir of {\it quasiparticles} which are then
allowed to tunnel. However, any experimental setup must consist of
{\it electrons} tunneling in and out of one or several reservoirs into
a FQH liquid. The edge excitations of the FQH liquid by themselves
cannot equilibrate since the liquid by itself does not dissipate and
there is no loss of phase coherence. The real loss of phase coherence
(and the resulting dissipation) originates in the electron reservoir,
{\it i.~e.\/} in the external leads. Hence, the question to be
addressed is how the tunneling of {\it electrons} into a FQH system is
affected by the nature of the quasiparticles of the FQH liquid and by
the equilibration mechanism. It is then natural to ask for what class
of contacts ({\it i.~e.\/} couplings with external leads) it is
possible to use tunneling experiments to reveal the nature of the FQH
quasiparticles. This is the main motivation of this work.

The purpose of this paper is thus to develop a conceptual framework
for the study of the non-perturbative physics of tunneling into a
chiral Luttinger liquid from a reservoir.  We will be interested on
the physics of these systems at general voltages and temperatures or,
equivalently, in the crossover from weak to strong coupling. For the
sake of simplicity, in this paper we discuss the problem of tunneling
to single edge QH liquids.  Recently, Kane and
Fisher\cite{Kane&Fisher3} described the behavior of a junction between
a QH state and reservoirs in the weak coupling regime where they found
that the two-terminal conductance is {\it not} universal.  In this
paper we will show that in the strong coupling limit, the two-terminal
conductance through a FQH state with filling fraction $\nu$ becomes
{\it universal}. This is the regime we will be primarily interested
in.  In particular, we will show that different ways of coupling to a
principal FQH state with $\nu=\frac{1}{2m+1}$ can lead to values of
the two-terminal conductances such as $G=\nu e^2/h$ (the usual one),
$G=\frac{2\nu}{\nu+1}e^2/h$ or simply $G=e^2/h$.  The main tool that
we will use in this work is a mapping (that we present below) of the
problem of tunneling from a higher dimensional electron gas to the
edges of a QH liquid to an {\it exactly} equivalent problem of
tunneling from a single-channel chiral Fermi liquid and the same QH
edge.  This is applicable if the coupling is made via a single
point-like contact. The injected electrons in the QH edge states, as
we show, are {\it not} in equilibrium with the reservoirs, and this is
the origin of the renormalized conductance. Equilibration is recovered
if the coupling to the QH state is made through many contacts. In the
case of many contacts we obtain an $I-V$ characteristic and compare it
with recent experimental data by A.~Chang {\it et. al.} for the full
range of probed voltages. This mapping can also be used to describe
the problem of tunneling of electrons from external leads to a
non-chiral Luttinger liquid\cite{Maslov&Stone}.

This paper is organized as follows. In section \ref{sec:sum} we
summarize the basic physical picture and results. In section
\ref{sec:oneimp} we show that the tunneling from a generic electron
gas to a FQH liquid through a point contact is equivalent to tunneling
from an effective {\it chiral } Fermi liquid to the edges of the same
FQH state. Using this mapping we solve non-perturbatively several
cases of interest, showing that the two-terminal conductance can
assume distinct universal values depending on the nature of the
contacts.  In section \ref{sec:one-half} we show that the $\nu=1$ to
$\nu=1/3$ QH junction is an experimental realization of the problem of
tunneling between two $g=1/2$ chiral Luttinger liquids. This is an
important theoretical system since it is exactly solvable for all
correlation functions. In section \ref{sec:many} we discuss the role of
multiple tunneling processes in the equilibration of QH edge states.
In this section we derive a universal crossover function for the
conductance of an electron gas to a QH edge state for arbitrary
voltages and temperature through a large number of individually weak
tunneling junctions. Since the number of junctions is large, the
effective coupling is large and this system cannot be described
perturbatively. The crossover function has a remarkable resemblance to
a Kondo-like effect.  At large voltages the conductance saturates at
$G=\nu {\frac{e^2}{h}}$.  We use this crossover function to fit
the experimental data of A.~Chang {\it et.~al.} over the entire range of
voltages\cite{Chang}. We also give a set of asymptotic values of the
universal conductance in the strong coupling limit for a
generalized problem of $N_L ,N_R$ contacts to left and right
reservoirs. Section \ref{sec:conc} is devoted to the conclusions. In
the Appendix we present the details of the mapping of a high
dimensional electron gas to a one-dimensional chiral Fermi liquid for
tunneling through a single impurity.

\section{Summary of the approach and results}
\label{sec:sum}

In this paper we show that for the purposes of studying tunneling
between an electron gas and a chiral Luttinger liquid, the electron
gas behaves as an effective chiral Fermi liquid. The idea, although
simple, is strongly justified for reasons other than simply looking
at the algebraic decay of the electron Green's function. Just as in
the Kondo problem, scattering through a point-like impurity potential
can be understood in terms of the radial motion of an effective
one-dimensional non-chiral fermion on a semi-infinite 
line (see for example Ref.~\cite{Affleck&Ludwig}). 
By unfolding the line around the impurity,
one obtains a chiral fermion on an infinite line.  Thus, it is well
grounded to expect that, for the problem of tunneling through a
single point, the electron gas that makes up the reservoir should
behave exactly in the same way as a chiral Fermi liquid or the edge
of a sharp $\nu=1$ QH state. This mode selection works regardless of
the number of spatial dimensions of the electron gas in the
reservoir.  One can think of this construction as the cleanest,
easiest, let alone cheapest, (virtual) realization of a sharp edge of
a $\nu=1$ QH state ($g=1$)!

We consider in detail the case of tunneling between an electron gas
and a QH liquid belonging to the principal sequence (Laughlin)
$\nu=\frac{1}{2m+1}$, $m$ integer. As noted above, the electron gas
can be regarded as a chiral Fermi liquid, so we have at hand the
problem of studying tunneling between Luttinger liquids with different
$g$. We perform a change of basis in order to rescale the radii of the
bosonic fields describing the chiral branches and in this way we map
the problem of tunneling from an electron gas to the edges of a
Laughlin FQH liquid with filling fraction $\nu$ to the problem of
tunneling between two identical chiral Luttinger liquids with
$g=\frac{2\nu}{\nu+1}$.

Using this framework, we point to new experimentally accessible
features of quantum Hall liquids which are consequences of the chiral
Luttinger liquid nature of the edge states. For example, we show that
in a two-terminal conductance experiment where one of the contacts is
a point-like tunneling center between a FQH liquid and an electron gas
(reservoir), the conductance in the strong coupling limit is larger
than the Hall conductance, but universally related to it. More
explicitly, the two-terminal conductance for strong tunneling (or high
voltages) to a Laughlin FQH state of filling fraction $\nu$ is
$G=\frac{2\nu}{\nu+1}e^2/h$. This implies that the electrons injected
into the FQH liquid will be hotter ({\it i.~e.}, higher voltage) than
those in the reservoir. The tunneling contact is thus a springboard
for raising the chemical potential\cite{foot1}.

Since the injected electron is hotter than the reservoir, one might
ask how to restore an equilibrium between the QH liquid and the
reservoir. We address this problem by considering the effects of many
impurities. It was first pointed out by Kane and Fisher that the
equilibration mechanism between QH liquids and reservoirs could be
understood in terms of tunneling through a line of weak coupling
impurities, which would bring the voltage along the edge states
monotonically to equilibrium with the reservoirs
\cite{Kane&Fisher3}. In this paper we show that the same is true in
the case of strong coupling impurities. The only difference between
these two situations is that, as the number of impurities increases,
the voltage of the QH edge oscillates around the asymptotic
equilibrium value.  One consequence of the equilibration with the
reservoirs is that the two-terminal conductance acquires the value of
the bulk Hall conductance.

\begin{figure}
\vspace{0cm}
\noindent
\hspace{.375 in}
\epsfxsize=2.4in
\epsfbox{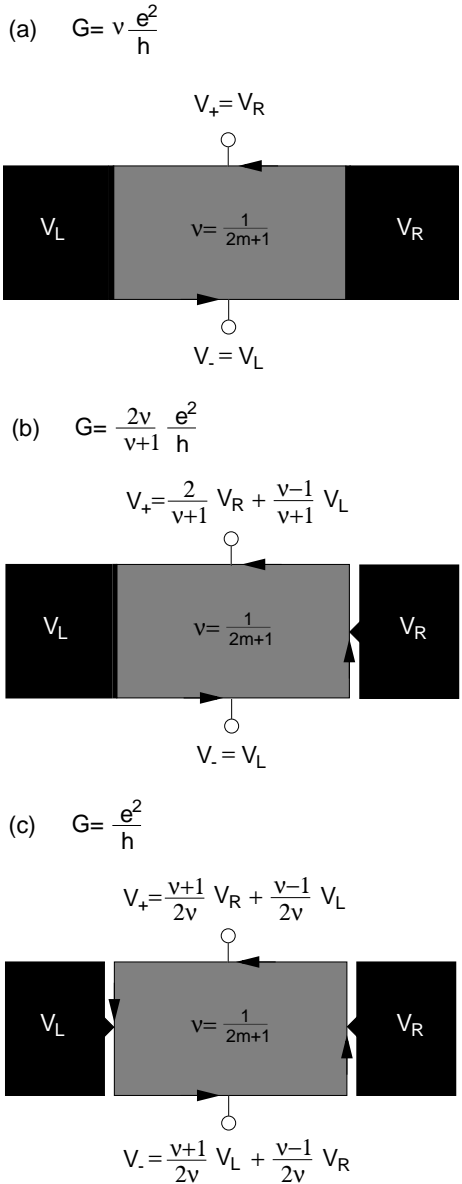}
\vspace{.2cm}
\caption{Different ways of strongly coupling a FQH liquid to
reservoirs gives distinct universal values for the two-terminal
conductance. In (a), the FQH liquid is coupled to the reservoirs via
many contacts. The edges are in equilibrium with their respective
reservoirs of departure, and the two-terminal conductance equals the
Hall conductance $\nu e^2/h$. In (b), the left reservoir is coupled
through many points, whereas the right one is coupled via a single
quantum channel (one point-like contact or impurity). The electrons at
the upper edge ($V_+$) are {\it not} in equilibrium with the right
reservoir ($V_R$). The two-terminal conductance is
$\frac{2\nu}{\nu+1}e^2/h$, {\it larger} than $\nu e^2/h$, and the Hall
voltage $V_+-V_-=\frac{2}{\nu+1}(V_R-V_L)>V_R-V_L$. In (c), both
reservoirs are coupled via a single point-like contact. Neither branch
is in equilibrium with either reservoir, and the two-terminal
conductance is $e^2/h$, independent of the filling fraction
$\nu=1/(2m+1)$.}
\label{fig1}
\end{figure}

The dependence of the two-terminal conductance on the way the
reservoirs are coupled to the FQH liquid is illustrated in
Fig. \ref{fig1}. In Fig. \ref{fig1}a we show the usual case where the
two-terminal conductance equals the Hall conductance $G=\nu e^2/h$. In
this case, the Hall voltage $V_+ -V_-=V_R-V_L$, and the edges are in
equilibrium with their respective reservoirs of departure. The
contacts are made through many points in this case.  In
Fig. \ref{fig1}b one reservoir is coupled through many points, whereas
the other is coupled through one single quantum channel, or point-like
contact. In this case, one edge branch is in equilibrium with one
reservoir ($V_-=V_L$), but the other is not ($V_+\ne V_R$). The
two-terminal conductance in this case is $G=\frac{2\nu}{\nu+1} e^2/h$,
{\it larger} than the Hall conductance, and thus the Hall voltage is
{\it higher} than the two-terminal voltage difference ($V_+
-V_-=\frac{2}{\nu+1}(V_R-V_L)$). Finally, in Fig. \ref{fig1}c both
contacts to the QH state are point-like, and neither edge branch is in
equilibrium with either of the two reservoirs. The two-terminal
conductance is $G=e^2/h$, again {\it larger} than the Hall
conductance, and the Hall voltage is also {\it higher} than the
two-terminal one ($V_+ -V_-=\nu^{-1}(V_R-V_L)$). In this last case,
there is an analogy to the problem of unrenormalized conductances in
quantum wires; there is one quantum channel going in and out of the
strongly correlated QH state, so that the conductance is $e^2/h$
regardless of the filling fraction $\nu=\frac{1}{2m+1}$.

The devices depicted in Fig \ref{fig1} are particular cases of the
general problem of connecting a Laughlin FQH liquid with $N_L$
point-contacts to the left reservoir, and $N_R$ point-contacts to the
right one. For strong coupling (which can always be achieved for large
enough voltages), we find that the universal conductance $G_{N_L,N_R}$
is given by
\begin{equation}
G_{N_L,N_R}=
\frac{\left[1-\left(\frac{\nu-1}{\nu+1}\right)^{N_L}\right]
\left[1-\left(\frac{\nu-1}{\nu+1}\right)^{N_R}\right]}
{\left[1-\left(\frac{\nu-1}{\nu+1}\right)^{N_L+N_R}\right]}
\ \nu\frac{e^2}{h}\ ,
\end{equation}
which reproduces the cases in Fig. \ref{fig1}: $G_{\infty,\infty}=\nu
e^2/h$, $G_{\infty,1}=\frac{2\nu}{\nu+1} e^2/h$, and $G_{1,1}= e^2/h$.

In this paper we also consider in detail the problem of {\it
equilibration} in the presence of many impurities or contacts coupling
a reservoir to the QH liquid. The case of many impurities is treated
in the following way. The propagation in the reservoir side between
spatially separated impurities takes place incoherently. Moreover,
after scattering from a point-like contact to the QH liquid, the
energy of the scattered electron would drop on the chiral Fermi
branch, but would then be brought back to equilibrium with the
reservoir. Thus, electrons incident into any of the impurities from
the reservoir side will always be at the same voltage. On the QH
liquid side, however, the energy of the scattered electron is
maintained from one scattering event to another, because of the
dissipationless nature of the QH state. Multi-impurity
scattering will bring, eventually, the QH edge to equilibrium with the
reservoir. This scattering mechanism allows one to obtain the solution
of the many impurity problem from the single impurity one. Notice that
the behavior of the electrons in the reservoir side, namely loss of
phase memory and equilibration, is what provides the simplification
needed in solving the multi-impurity case. Clearly, incoherence is a
key assumption here.  It must hold for well separated impurities
which should act essentially independently from each other. This
situation is familiar from the physics of dilute magnetic alloys where
the Kondo impurities are indeed independent from each other. It is
also clear that as the impurities ({\it i.~e.} the contacts) become
closer to each other, coherence effects in the reservoir should become
important. In this limit, we expect a richer and more complex quantum
behavior much like the multi-impurity or multichannel behavior in
Kondo systems.

We use this framework for the multi-impurity case and apply it to
obtain a solution for the case of tunneling, through many impurities,
from an electron gas to a $\nu=1/3$ liquid, which is of direct
relevance for comparison to the experiments of A. Chang {\it et.~ al.}
\cite{Chang}.
We show that the mechanism that we consider in this work contains the
necessary ingredients to explain not only the low voltage, low
temperature power law behavior, but also the breakdown voltage scale
for deviations from the anomalous power law scaling.  In addition, it
predicts the asymptotic large voltage conductances, which should
saturate at the bulk value of the Hall conductance $\nu
{\frac{e^2}{h}}$.  We show that the conductance between the electron
gas and the $\nu=1/3$ FQH state is given by
\begin{equation}
G=\nu\frac{e^2}{h}
\left\{
1-\frac{e^{-\frac{1}{2}\left(\frac{2\pi T}{T_K}\right)^2}}
{\left[\left(1-e^{-\left(\frac{2\pi T}{T_K}\right)^2}\right)
\left(\frac{V}{2\pi T}\right)^2+1
\right]^{3/2}}\right\}\ ,
\label{fullconductanceintro}
\end{equation}
where $T$ is the temperature and $T_K$ is a crossover energy scale
determined by the couplings of {\it all} impurities connecting the
electron gas to the QH state. The only assumption made in the
derivation of Eq. (\ref{fullconductanceintro}) is that individual
impurities are weakly coupled.  This assumption, as we show, does not
restrict the result to the regime of small conductances since the
scale $T_K$ can actually take a broad range of values. Hence the
expression should remain valid even for voltages large compared to
$T_K$ and conductances as large as the bulk Hall value $\nu
{\frac{e^2}{h}}$ (this point is made clear below).

The result of Eq. (\ref{fullconductanceintro}) can be used for
comparison with the experimental data. One can easily check that, for
$V\ll T_K$, Eq. (\ref{fullconductanceintro}) reproduces the scaling
form used in Ref. \cite{Chang}. The voltage scale for which the
experimental data departs from the low voltage scaling form is
determined by the energy scale $T_K$, which is evident in
Eq.(\ref{fullconductanceintro}). This breakdown voltage scale can be
determined from the low voltage data, since the amplitude of the
tunneling conductance is directly related to $T_K$ and $T$:
\begin{equation}
G_0=\lim_{V\to 0}\ G=\nu\frac{e^2}{h}
\left[1-e^{-\frac{1}{2}\left(\frac{2\pi T}{T_K}\right)^2}\right]\ .
\end{equation}
Therefore, the ``breakdown'' voltage is indeed part of the full
solution to the problem of tunneling into the FQH edge.

One should not overlook the fact that the experimentally measured
conductance saturates to the Hall value $\nu {\frac{e^2}{h}}$. This is
a signature of many impurities, and it is in agreement with
Eq. (\ref{fullconductanceintro}). We contrast this case to the one
impurity contact (to one of the reservoirs, as in Fig. \ref{fig1}b),
for which the two-terminal conductance should reach the value
$G=\frac{2\nu}{\nu+1}{\frac{e^2}{h}}$, and electrons enter the edge of
the QH liquid hotter than the reservoir.

Finally, we would like to point out that, as a natural consequence of
the physics of the point-contact junctions between bulk electron gases
and FQH states that are discussed here, these junctions constitute a
simple physical realization of the DC voltage transformer proposed
very recently by Chklovskii and Halperin\cite{Chklovskii&Halperin}.
The point-contact junctions are more readily realizable and possibly
avoid many of the difficulties discussed by Chklovskii and Halperin.

\section{Tunneling from an electron gas to a FQH edge}
\label{sec:oneimp}

Our starting point is the Lagrangian density that describes the
dynamics on the edge of the FQH liquid, the electron gas reservoirs,
and the tunneling between the two via a single impurity:
\begin{equation}
{\cal L}={\cal L}_{\rm edge}+{\cal L}_{\rm gas}+
{\cal L}_{\rm tun}\ .
\end{equation}
The edge excitations of a FQH liquid with $\nu=\frac{1}{2m+1}$ are
described by a single free chiral boson field $\phi$ with 
\begin{equation}
{\cal
L}_{\rm edge}={1\over 4\pi}\partial_x\phi (\partial_t- v
\partial_x)\phi
\end{equation}
and equal-time commutation relation $[\phi(t,x),\phi(t,y)]=-i\pi\ {\rm
sgn}(x-y)$. The edge electron operator is written in terms of the
boson field as $\psi_{\rm edge}(t,x)\propto
e^{-i\frac{1}{\sqrt{\nu}}\phi(t,x)}\ $ \cite{XGWcll}. ${\cal L}_{\rm
gas}$ is the Lagrangian for the electron gas, with $\psi(t,x)$ the
electron operator there. The tunneling Lagrangian is then
\begin{equation}
{\cal L}_{\rm tun}=\sqrt{2\pi}\Gamma\ \delta(x)\ e^{i\omega_J t} \ \
e^{i\frac{1}{\sqrt{\nu}}\phi(t,0)}\psi(t,0)+H.c.\ \ .
\end{equation}
The Josephson frequency $\omega_J=eV/\hbar$ is set by the difference
of voltages between the electron gas and the {\it incoming} edge
branch to the impurity. As we will see later, the voltage for the {\it
outgoing} branch is raised due to tunneling.

The next step is the mapping between a 2D or 3D electron gas to a 1D
chiral Fermi liquid for tunneling through a single impurity. We
follow the conventional procedure used in the theory of the Kondo problem
(see, for instance, Ref.~\cite{Affleck&Ludwig}) where it is shown that 
the problem of a single isotropic magnetic impurity coupled to
a three-dimensional Fermi liquid is mapped to the problem of a 
one-dimensional chiral Fermi liquid coupled to the impurity. 
The same conclusions hold for any
number of dimensions greater than one. In the Kondo problem, one
basically writes the electron wavefunction $\psi(\vec{r})$ in a
spherically symmetric basis, with the origin at the impurity position
$\vec{r}=0$. The operator $\psi(\vec{0})$ depends only on the $L=0$
harmonic, so that this is the only channel that participates in
coupling to the impurity. For the case of two dimensions, only the
$m=0$ channel is coupled.

One may worry that if the impurity is on the planar boundary of the 3D
electron gas, the spherical symmetry is spoiled. However, the
spherical symmetry is not a necessary condition to arrive at the
conclusion that only one quantum channel is coupled via the impurity.
We show in appendix \ref{appendiximp}, for a very general set of
problems, that one can always find a basis of eigenstates of the
Hamiltonian for the electron gas such that only one quantum channel
couples to the impurity, and again only the radial components of that
channel are important.  The electron operator for this channel can be
written in terms of left and right moving fermions on a half line,
which is the radial coordinate $r$ (left and right moving particles
correspond to incoming and outgoing particles with respect to the
impurity). By unfolding the half line into a full line, one can
describe the electron operators in terms of a single chiral fermion on
an infinite line. Therefore, for tunneling through a point-like
contact, the semi-infinite 3D electron gas becomes effectively
equivalent to a 1D chiral Fermi liquid. It can be regarded in much the
same way as if it were a sharp edge of $\nu=1$ QH state.

Thus, we can write
\begin{equation}
{\cal L}_{\rm gas}={\cal L}_{\rm gas}^{\bf \alpha}+
{\rm other\  channels}\ ,
\end{equation}
where ${\bf \alpha}$ labels the quantum numbers of the channel that
couples to the impurity (which depend on the symmetry in the problem,
such as $L=0$ for the Kondo problem). We can then bosonize the ${\bf
\alpha}$ effective chiral 1D mode corresponding to the (unfolded)
radial direction:
\begin{equation}
{\cal L}_{\rm gas}^{\bf \alpha}={1\over 4\pi}\partial_x\varphi
(\partial_t- {\tilde v} \partial_x)\varphi
\end{equation}
with the electron operator given (in terms of $\varphi$) by
$\psi(t,x)=\frac{1}{\sqrt{2\pi}}:e^{-i\varphi(t,x)}:\ $.

We are free to rescale the position coordinates and thus alter the
velocities. Moreover, because the tunneling takes place at a point,
and also the two chiral boson fields $\phi$ and $\varphi$ are in
separate spaces, we are free to rescale the position coordinates
independently, allowing us to set both $v=\tilde v=1$. (Notice that
even though we used the same symbol $x$ for the coordinates of both
fields, $x$ separately parametrizes the fields along their arc length.
The only commom point in the parametrization is $x=0$, which is the
impurity location.)

The tunneling problem is then described by the Lagrangian
\begin{eqnarray}
{\cal L}&=&
{1\over 4\pi}\partial_x\phi
(\partial_t-\partial_x)\phi
+{1\over 4\pi}\partial_x\varphi
(\partial_t-\partial_x)\varphi\label{Leff}\\
&\ &\ \ \ \ +\Gamma\ \delta(x)\ e^{i\omega_J t}
\ \ e^{i\left(\frac{1}{\sqrt{\nu}}\phi(t,0)-\varphi(t,0)\right)}+H.c.
\nonumber
\end{eqnarray}
We can bring the tunneling term to more familiar forms by means of a
rotation. Instead of doing this directly for Eq. (\ref{Leff}), let us
treat a more general problem. Consider tunneling between a
$\nu_1=1/n_1$ and a $\nu_2=1/n_2$ QH state, with both $n_{1,2}$ odd
integers (single edges). The tunneling coupling is
\begin{eqnarray}
{\cal L}_{\rm tun}&=&
\Gamma\ \delta(x)\ e^{i\omega_J t}
\ \ e^{i\left(\frac{1}{\sqrt{\nu_1}}\phi_1(t,0)-
\frac{1}{\sqrt{\nu_2}}\phi_2(t,0)\right)}+H.c.\\
&=&
\Gamma\ \delta(x)\ e^{i\omega_J t}
\ \ e^{i\frac{1}{\sqrt{\bar\nu}}\left(\phi_a(t,0)-
\phi_b(t,0)\right)}+H.c.
\end{eqnarray}
where we have performed the $O(2)$ rotation
\begin{equation}
\left(\matrix{\phi_a\cr \phi_b}\right)
=
\left(\matrix{
\ \cos\theta & \sin\theta\cr
-\sin\theta & \cos\theta}\right)
\left(\matrix{
\phi_1\cr \phi_2}\right)\ ,
\end{equation}
with
\begin{eqnarray}
\cos\theta&=&\frac{1}{\sqrt{2}}
\frac{\sqrt{\nu_1^{-1}}+\sqrt{\nu_2^{-1}}}
{\sqrt{\nu_1^{-1}+\nu_2^{-1}}}\nonumber\\
\sin\theta&=&\frac{1}{\sqrt{2}}
\frac{\sqrt{\nu_1^{-1}}-\sqrt{\nu_2^{-1}}}
{\sqrt{\nu_1^{-1}+\nu_2^{-1}}}\nonumber\ .
\end{eqnarray}
The tunneling term corresponds to tunneling between two chiral
Luttinger liquids with
\begin{equation}
g=\bar\nu^{-1}=\frac{\nu_1^{-1}+\nu_2^{-1}}{2}\ .
\end{equation}
In particular, since we showed that the electron gas couples to a
single impurity through only one quantum channel (see Appendix
\ref{appendiximp}), it is effectively equivalent to a 1D chiral Fermi
liquid (virtual $\nu=1$), and so we have
\begin{equation}
g=\frac{\nu^{-1}+1}{2}\ ,
\end{equation}
where $\nu$ is the filling fraction of the QH state coupled to the
electron gas (reservoir) through the single impurity.

\begin{figure}
\vspace{.8cm}
\noindent
\epsfxsize=3in
\epsfbox{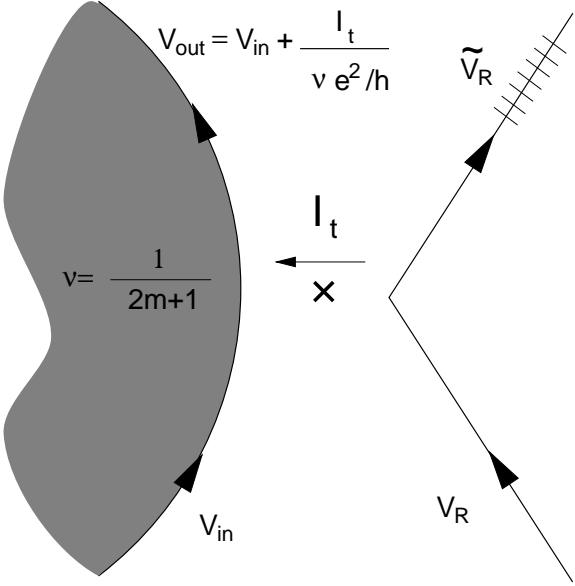}
\vspace{.8cm}
\caption{One impurity scattering picture. The only voltages that
affect the tunneling current are the {\it incoming} ones, $V_{in}$ and
the reservoir voltage $V_R$. The outgoing voltage $V_{out}$ is raised
due to the injected current $I_t$. The outgoing voltage on the
reservoir side ${\tilde V}_R$ also depends on the tunneling current;
however, after scattering, electrons in this branch are equilibrated
again with the reservoir.}
\label{fig2}
\end{figure}

We can then use the known results for tunneling between two Luttinger
liquids and apply to the problem of tunneling from an electron gas to
the edge of a FQH liquid. We can, for example, determine the current
injected into the edge branch, which depends on the voltage difference
between the electron gas and the {\it incoming} branch to the
impurity, as shown in Fig. \ref{fig2}.

Because of the weak-strong coupling duality symmetry present in the
problem of tunneling between Luttinger liquids, we can also turn to
the dual picture corresponding to tunneling between two Luttinger
liquids with
\begin{equation}
\tilde g=\frac{1}{g}=\frac{2}{\nu^{-1}+1}\ .
\end{equation}

The differential tunneling conductance depends on both $g$ (or
${\tilde g}$) and $V_R-V_{in}$. At zero temperature, it is given by
\cite{Fendley1,Weiss}
\begin{equation}
G_t={\tilde g}\ \frac{e^2}{h}
\times
\cases{
\sum_{n=1}^{\infty}c_n(1/{\tilde g})
\ \left(\frac{V}{2T_K}\right)^{2n(1/{\tilde g}-1)},
&\mbox{$\frac{V}{2T_K}< e^{\delta}$}\cr
1-\sum_{n=1}^{\infty}c_n({\tilde g})
\ \left(\frac{V}{2T_K}\right)^{2n({\tilde g}-1)},
&\mbox{$\frac{V}{2T_K}> e^{\delta}$}\cr
}
\label{Gexpansion}
\end{equation}
where $V$ is the voltage difference between the reservoir and the
incoming edge branch, $V_R-V_{in}$. The $T_K$ is an energy scale set
by the tunneling amplitude $\Gamma$ ($T_K\propto |\Gamma|^{-1/
({\tilde g}^{-1}-1)}$), and the coefficients $c_n$ are given by
\begin{equation}
c_n({\tilde g})=(-1)^{n-1}
\ \frac{\Gamma(n{\tilde g}+1)}{\Gamma(n+1)}
\ \frac{\Gamma(1/2)}{\Gamma(n({\tilde g}-1)+1/2)}\ .
\end{equation}
The domains of convergence of the dual series are restricted by
$\delta=[{\tilde g}\ln {\tilde g}+(1- {\tilde g})\ln (1-{\tilde
g})]/[2({\tilde g}-1)]$

Due to the injected current, the voltage of the edge branch past the
impurity is raised by $\Delta V=I_t/(\nu e^2/h)$.

We now turn to the case of strong coupling ($|\Gamma|\to \infty$,
$T_K\to 0$), or equivalently, large voltage differences ($V\gg
T_K$). In this case, the tunneling conductance reaches the strong
coupling asymptotic value $G_t={\tilde g}e^2/h$. We then apply this
result to the geometries depicted in Fig. \ref{fig1}.

\subsection{One point-like contact}
This is the case shown in Fig. \ref{fig1}b. One edge branch is in
equilibrium with its reservoir of departure ($V_-=V_L$), because this
contact is made through many impurities (see section
\ref{sec:many}). For the other contact, the value of the voltage level
$V_+$ depends on the injected current from the right reservoir. The
injected current for strong coupling is given by
\begin{equation}
I_t={\tilde g}\  \frac{e^2}{h}\ (V_R-V_-)
=\frac{2}{\nu^{-1}+1}\frac{e^2}{h}\ (V_R-V_-)\ ,
\end{equation}
and consequently
\begin{equation}
V_+-V_-=\nu^{-1}{\tilde g}\ (V_R-V_-)
=\frac{2}{\nu+1}\ (V_R-V_-)\ .
\end{equation}
Thus, the voltage $V_+$ is given by
\begin{equation}
V_+=\frac{2}{\nu+1}V_R\ +\ \frac{\nu-1}{\nu+1}V_L\ ,
\end{equation}
and the Hall voltage is
\begin{equation}
V_+-V_-=\frac{2}{\nu+1}\ (V_R-V_L)\ .
\end{equation}

The two-terminal conductance of the device (determined from
$I_t/(V_R-V_L)$) is
\begin{equation}
G=\frac{2\nu}{\nu+1}\frac{e^2}{h}\ .
\end{equation}

\subsection{Two point-like contacts}
This is the case shown in Fig. \ref{fig1}c. Neither of the edge
branches is in equilibrium with either reservoir. We have to determine
the voltages from the tunneling currents in both contacts. The voltages
$V_+$ and $V_-$ are obtained from
\begin{eqnarray}
V_+-V_-&=&\nu^{-1}{\tilde g}\ (V_R-V_-)=
\frac{2}{\nu+1}\ (V_R-V_-) \nonumber\\
V_--V_+&=&\nu^{-1}{\tilde g}\ (V_L-V_+)=
\frac{2}{\nu+1}\ (V_L-V_+) \ ,\nonumber
\end{eqnarray}
which give
\begin{eqnarray}
V_+&=&\frac{\nu+1}{2\nu}V_R+\frac{\nu-1}{2\nu}V_L \nonumber\\
V_-&=&\frac{\nu+1}{2\nu}V_L+\frac{\nu-1}{2\nu}V_R \ .\nonumber
\end{eqnarray}
The Hall voltage is given by
\begin{equation}
V_+-V_-=\nu^{-1}\ (V_R-V_L)\ .
\end{equation}

The current flowing through the device is
\begin{equation}
I_t=\nu \frac{e^2}{h}\ (V_+-V_-)=\frac{e^2}{h}\ (V_R-V_L)\ ,
\end{equation}
and therefore the two-terminal conductance is
\begin{equation}
G=\frac{e^2}{h}\ .
\end{equation}

\begin{figure}
\vspace{.8cm}
\noindent
\hspace{.125 in}
\epsfxsize=3in
\epsfbox{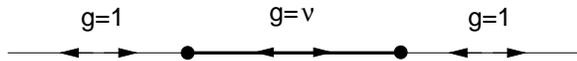}
\vspace{.8cm}
\caption{If the contacts from the FQH liquid to the reservoirs are
made through a single point each, the edges can be deformed so as to
line them up much in the same way as in a non-chiral Luttinger liquid
(quantum wire), and so $G=e^2/h$. However, in contrast to quantum
wires, the ``extra'' spatial dimension in the case of the FQH state,
allows measurements (such as the ratio $\nu^{-1}$ between the Hall
voltage to the two-terminal voltage), which probes the Luttinger
liquid nature of the state in between the reservoirs.}
\label{fignoncll}
\end{figure}

This result is the universal conductance for spinless non-interacting
electrons. It resembles the result for the conductance of quantum
wires, where the reservoirs mask the (non-chiral) Luttinger liquid
behavior of the 1D wire \cite{Tarucha,Maslov&Stone}. We can understand
why there is such correspondence in a very simple way. Because the
contacts are made to the two reservoirs at a single point each, we can
deform the edges of the QH state so as to bring them to line up on top
of each other at a segment that connects the two contacts.
The end result is that the two opposite chiralities
overlap on the same segment (see Fig. \ref{fignoncll}), which makes
the problem the same as its non-chiral counterpart of
Ref. \cite{Tarucha,Maslov&Stone}. Notice also that we have shown that,
for point-like contacts to the reservoirs, the electron gas making up
the external lead behaves
effectively as a 1D chiral Fermi liquid, which was the model for the
leads that Maslov and Stone proposed recently\cite {Maslov&Stone}.
In their picture, the non-chiral Luttinger liquid of a quantum wire
turned adiabatically into a Fermi liquid (describing the leads) which
they also took to be a one-dimensional Fermi system. Our mapping shows
that only one mode of the electrons in the reservoir will effectively 
tunnel into the Luttinger liquid. Furthermore, even if the coupling
between the leads and the wire is very smooth, there are always 
backscattering processes at the crossover. These processes are always
relevant operators and the system always flows to strong coupling where
the tunneling picture is accurate\cite{Kane&Fisher1}. 
Thus, the mapping presented here
provides a formal justification of the model of Maslov and Stone.

Having mentioned the similarities, let us now look at the {\it
important} differences between transmitting current in quantum wires
and FQH liquids. The chiral 1D Luttinger liquid in the case of the FQHE
comes from the edges of the 2D strongly correlated FQH liquid, and we
can explore the extra one dimension experimentally. For example, one
can measure voltages at the two edges ($V_+$ and $V_-$ in
Fig.~\ref{fig1}c) because they are spatially separated, and by doing
so probe Luttinger liquid behavior between the reservoirs. This is not
possible in the case of quantum wires, for the two chiralities are
mixed and one cannot measure the two chemical potentials separately
inside the wire. A measurement of the Hall voltage
$V_+-V_-=\nu^{-1}(V_R-V_L)$ returns a value that is {\it larger} than
the two-terminal voltage ($V_R-V_L$) by a factor of $\nu^{-1}$; this
factor contains information on the Luttinger liquid behavior between
the reservoirs which is experimentally measurable, and has no
counterpart in quantum wires.

Another important difference between quantum wires and devices such as
the ones we discuss in this paper is that quantum wires can be
connected to leads in only one way ({\it i.~e.}, one end to each lead), 
whereas we can
connect a FQH state (a 2D system) to, say, $N_L$ contacts to one
reservoir and $N_R$ to the other. The consequence is that in quantum
wires one can only observe a conductance $e^2/h$ (and multiples of
it), whereas for FQH states the conductance $G_{N_L,N_R}$, as we show
in section \ref{sec:many}, can take values in whole families of
rational multiples of $e^2/h$.  (Examples are the cases shown in
Figs. \ref{fig1}b and \ref{fig1}c, which correspond to 
$G_{\infty,\infty}=\nu
e^2/h$ and $G_{\infty,1}=\frac{2\nu}{\nu+1}e^2/h$, respectively.)

In summary, the strong coupling (large voltage) value of the tunneling 
conductance from a reservoir ( {\it i.~e.~}, a
Fermi liquid) to the chiral Luttinger liquids of the edges of a FQH
state, in spite of being universal,
depends not only on the properties of the FQH liquid but also on the
nature of the contacts. Moreover, the physics of the edge states of 
the FQH states can be probed with a variety of junctions for which there is 
no counterpart in one-dimensional quantum wires.

\section{Experimental realization of g=1/2}
\label{sec:one-half}
A very interesting case is when $\nu=1/3$, in which case $g=2$. This
case is the dual point to ${\tilde g}=1/2$, which is exactly solvable,
not only for the conductance, but for the full noise spectrum (indeed,
{\it all} $n$-point correlation functions). This non-trivial exactly
solvable point is very important theoretically because it provides the
comparison ground for any result obtained perturbatively.  Before, it
was simply a theoretical tool, with no physical realization. The
tunneling from an electron gas to a $\nu=1/3$ state makes it now
possible to study the important $g=1/2$ state experimentally.

Let us focus on this $g=1/2$ case and a single impurity. The exact
solution (see Refs.\cite{Kane&Fisher1,Fendley1,CFW1} and references
therein) for the tunneling current and differential conductance is
\begin{eqnarray}
I_t&=&e\int\frac{d\omega}{2\pi}\ \frac{\omega^2}{\omega^2+(T_K/2)^2}
\ \left(f(\omega-\omega_0)-f(\omega)\right)\\
G_t&=&\frac{1}{2}\frac{e^2}{h}
\int d\omega\ \frac{\omega^2}{\omega^2+(T_K/2)^2}
\ \left(-f'(\omega-\omega_0)\right)\ ,
\end{eqnarray}
where $f=(1+e^{\omega/T})^{-1}$ is the Fermi distribution,
$\omega_0=\frac{1}{2}\frac{eV}{\hbar}$, and
$T_K=(2\pi|\Gamma|)^{-1}$. The $\omega_0$ and $T_K$ are,
besides the temperature, the two energy scales in the problem. As one
varies the voltage, the current and conductance scaling with $V$
changes depending on the relative value of $V$ as compared to $T$ and
$T_K$. Consider, for example, the case when $T\ll T_K$. In this
case, the conductance scales as
\begin{equation}
G_t\approx\ \frac{1}{2}\frac{e^2}{h}
\cases{
\frac{1}{3}\left(\frac{2\pi T}{T_K}\right)^2,&\mbox{$V\ll T$}\cr
\ \ \left(\frac{V}{T_K}\right)^2\ \ \ ,&\mbox{$T\ll V\ll T_K$}\cr
\ \ \ \ 1\ \ \ \ \ \ \ ,&\mbox{$V\gg T_K$}\cr
}\ .
\end{equation}
The anomalous scaling of the current with voltage that is
characteristic of Luttinger liquids takes place between the two energy
scales $T$ and $T_K$. In recent experiments, Chang {\it et. al.}
have observed such scaling. In the experimental data, the $G_t\propto
V^\alpha$ power law scaling breaks down at a certain voltage. One is
inclined to think that $T_K$ provides the energy scale for the
breakdown, although in that particular experiment tunneling takes
place not through one, but many impurities. The idea that there is
another energy scale in the problem, however, is still an important
one, and we will show that indeed such a breakdown scale also occur
for the many impurity problem.

\section{Multiple impurities and Equilibration}
\label{sec:many}

We will here show how to treat the problem of tunneling from an
electron gas or reservoir to a QH liquid via many impurities. In the
previous section we considered scattering through a single impurity
(see Fig. \ref{fig2}), in which case the incident edge channel comes
at a voltage level $V_{in}$, and the electron from the reservoir at
$V_{R}$. After scattering, the edge is at $V_{out}$, and the reservoir
branch at ${\tilde V}_R$. We will obtain the many impurity result by
considering the single scattering event as building blocks, and
cascading them.

\begin{figure}
\vspace{.8cm}
\noindent
\hspace{.125 in}
\epsfxsize=3in
\epsfbox{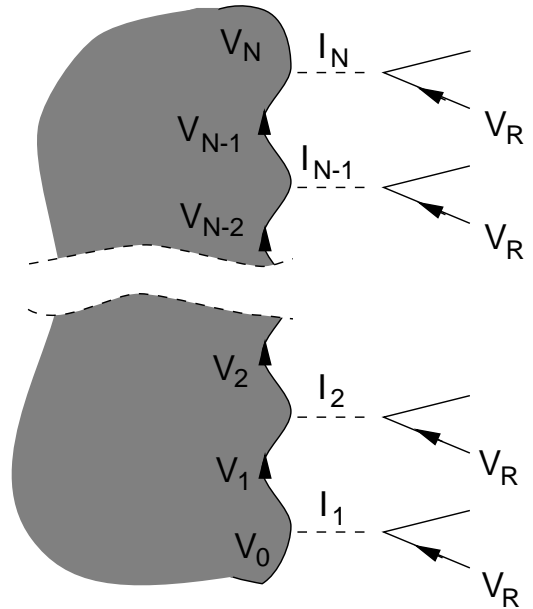}
\vspace{.8cm}
\caption{Multi ($N$) impurity scattering, assembled from the one
impurity building block. Notice that the voltages on the FQH liquid
side are maintained in between scattering events, whereas on the
reservoir side electrons come for the scattering always at $V_R$.}
\label{fig3}
\end{figure}

An important issue for this cascading is that of how the scattered
electrons on the reservoir side behave differently from those in the
QH side in between scattering events. In the reservoir side,
regardless of the voltage after scattering, the electrons will
equilibrate with the reservoir by the time of the next scattering
event. It will also lose its phase memory. This means that on the
reservoir side, electrons will always arrive for the scattering events
at the same voltage, namely, $V_{R}$. On the QH side, however, the
voltage is maintained between scattering events, and is
accumulated. Thus, the cascade can be assembled as shown in
Fig. \ref{fig3}.

The voltages past the $n$-th stage of the cascade are labeled $V_n$,
and the current flowing from the reservoir to the QH liquid is
$I_n$. The voltage difference between the outgoing ($V_n$) and
incoming ($V_{n-1}$) edge states is obtained from the current $I_n$
and the Hall conductance for the QH liquid:
\begin{equation}
V_n-V_{n-1}=\frac{I_n}{\nu \frac{e^2}{h}}\ .
\end{equation}
Now, the current $I_n$ is simply the single impurity tunneling
current, which depends on the incoming state ($n-1$) and the coupling
strength for the $n$-th impurity, as well as the temperature. The
coupling strength enters as an effective temperature $T_K^{(n)}$ for a
given impurity.

Notice, however, that the voltages $V_n$ alone are not sufficient to
describe the quantum states at the different stages of the cascade.
One needs the states $|\Phi_n\rangle=|V_n;\{n_k\}_n\rangle$, where the
$n_k$ are the occupation numbers of the oscillator modes of the chiral
boson $\phi$ describing the edge state. The voltage $V_n$ measures the
total charge or the zero mode of the boson field. The non-zero modes
should in principle affect the $I-V$ characteristics for each
impurity. An intuitive picture one can use to see the effects of the
excited oscillator modes is that they could account for an
``effective'' increase in the temperature (since a higher temperature
brings more excited modes). This ``effective'' increase in the
temperature enters in the expression for the tunneling current, in
addition to the voltages $V_n$ and $V_R$.

Nonetheless, there are two regimes where the oscillator modes are not
excited: weak and strong coupling. Effects due to the oscillator
modes, as described above, become only important at intermediate
coupling. For the weak and strong coupling cases we can simply use the
$I-V$ characteristics for a single impurity to obtain recursively the
voltages $V_n$ for all $n$, and we can obtain the total current that
flows to the QH liquid from $I=\nu \frac{e^2}{h}(V_N-V_0)$, where $N$
is the last impurity on the line, and $V_0$ is the initial voltage
level of the edge (in equilibrium with the other reservoir). The
recursion equation is
\begin{equation}
V_n-V_{n-1}=\frac{1}{\nu e^2/h}\ I_t(V_{R}-V_{n-1},T_K^{(n)},T)\ .
\label{recursion}
\end{equation}
Let us study solutions of these recursion relations for different
regimes of tunneling strengths.

\subsection{Weak coupling}

Without loss of generality, let us concentrate on the case
$g=2={\tilde g}^{-1}$. We will not use the fact that ${\tilde g}=1/2$
is exactly solvable; we just choose this example for clarity of
presentation, and because it is applicable to tunneling from an
electron gas to a $\nu =1/3$ FQH state. (We will present the result
for general $g$ after the derivation for $g=2={\tilde g}^{-1}$).

If the individual couplings are small, the $T_K^{(n)}$ will be large, in
which case we can use for the individual impurities the low voltage
($V\ll T_K^{(n)}$) expression for the current:
\begin{equation}
I_n=\frac{e^2}{h}\frac{T_K^{(n)}}{6}
\left(\frac{2\pi T}{T_K^{(n)}}\right)^3
\left[\frac{V_R-V_{n-1}}{2\pi T}+
\left(\frac{V_R-V_{n-1}}{2\pi T}\right)^3\right]\ .
\end{equation}
Now, we let $x_n=\frac{V_R-V_n}{2\pi T}$ and substitute it into the
recursion relation Eq. (\ref{recursion}), obtaining
\begin{equation}
x_n-x_{n-1}=-\frac{1}{6\nu }\left(\frac{2\pi T}{T_K^{(n)}}\right)^2
\ (x_{n-1}+x_{n-1}^3)\ ,
\label{recursion2}
\end{equation}
which we can transform into a differential equation, since the
couplings are assumed to be small:
\begin{equation}
\frac{dx}{dn}=-\frac{1}{6\nu }\left(\frac{2\pi T}{T_K^{(n)}}\right)^2
\ (x+x^3)\ ,
\end{equation}
to be integrated from the initial $x_0$ to the final $x_N$
(past the last impurity $N$) yielding
\begin{equation}
\int_{x_0}^{x_N}\frac{dx}{x+x^3}=-\sum_{n=1}^{N}\frac{1}{6\nu}
\left(\frac{2\pi T}{T_K^{(n)}}\right)^2=
-\frac{1}{2}\left(\frac{2\pi T}{T_K}\right)^2\ ,
\end{equation}
in which we define the effective $T_K$ from the individual
$T_K^{(n)}$. A very important point to be noticed is that one needs
not assume that the couplings $T_K^{(n)}$ are uniform; they can
fluctuate, and the only important parameter is the effective $T_K$
which incorporates even the fluctuations.
After integration, one obtains
\begin{equation}
\frac{x_N}{\sqrt{1+x_N^2}}=e^{-\frac{1}{2}\left(\frac{2\pi T}{T_K}\right)^2}
\frac{x_0}{\sqrt{1+x_0^2}}\ ,
\end{equation}
or equivalently,
\begin{equation}
V_N=V_R-2\pi T\left\{e^{\left(\frac{2\pi T}{T_K}\right)^2}
\left[1+\left(\frac{V}{2\pi T}\right)^{-2}\right]-1\right\}^{-1/2}\ ,
\end{equation}
where $V=V_R-V_L$. We have used that $x_0=\frac{V}{2\pi T}$, taking
the incoming edge at equilibrium with the other reservoir
($V_0=V_L$). The total current flowing to the QH liquid is obtained
from the voltage difference $V_N-V_0$, and is given by
$I=\nu\frac{e^2}{h}(V_N-V_0)$. The differential conductance then
yields

\end{multicols}
\widetext

\vspace{.5cm}
\noindent
\rule{.48\linewidth}{.3mm}\rule{.3mm}{.5cm}
\vspace{.5cm}

\begin{equation}
G=\nu\frac{e^2}{h}
\left\{
1-\frac{e^{-\frac{1}{2}\left(\frac{2\pi T}{T_K}\right)^2}}
{\left[\left(1-e^{-\left(\frac{2\pi T}{T_K}\right)^2}\right)
\left(\frac{V}{2\pi T}\right)^2+1
\right]^{3/2}}\right\}\ .
\label{fullconductance}
\end{equation}
The derivation for general $g$ is similar, and gives

\begin{equation}
G=\nu\frac{e^2}{h}
\left\{
1-\frac{e^{-\frac{1}{2}\left(\frac{2\pi T}{T_K}\right)^{2(g-1)}}}
{\left[\frac{1}{\Gamma^2(g)}
\left(1-e^{-(g-1)\left(\frac{2\pi T}{T_K}\right)^{2(g-1)}}\right)
\left(\frac{V}{2\pi T}\right)^{2(g-1)}+1
\right]^{\frac{2g-1}{2(g-1)}}}\right\}\ .
\label{fullotherg}
\end{equation}

\vspace{.5cm}
\noindent
\rule{.52\linewidth}{.0mm}\rule[-.47cm]{.3mm}{.5cm}\rule{.48\linewidth}{.3mm}
\vspace{.5cm}

\begin{multicols}{2}
\narrowtext

The result expressed in equation (\ref{fullconductance}) holds for all
values of $V,T,T_K$, as long as the assumptions used to derive the
expression holds, namely, that each individual impurity is weakly
coupled. Notice that, even though the $T_K^{(n)}$ are large, the
effective $T_K$ for the the line of impurities can be made small by
simply having a very long line (large $N$). Experimentally one has two
ways of varying $T_K$: changing the tunneling barriers, and taking
wider samples.

We can use Eq. (\ref{fullconductance}) for comparison with recent
experiments by A. Chang {\it et.al.} \cite{Chang}. The experiments
demonstrate power law scaling of the tunneling current with
voltage. For small voltages, the data is well fitted by a universal
$I-V$ characteristic which crosses over from a linear regime for $V\ll
T$, to the power law scaling for $V\gg T$. One can easily check that
the universal curve used for the conductance correspond to the $T,V\ll
T_K$ limit of Eq. (\ref{fullconductance}), namely
\begin{equation}
G\approx 
\nu\frac{e^2}{h}
\frac{1}{2}
\left(\frac{2\pi T}{T_K}\right)^2
\left[
1+3\left(\frac{V}{2\pi T}\right)^2\right]\ .
\label{smallVT}
\end{equation}
However, the fit used in Ref. \cite{Chang} breaks down beyond a
certain voltage scale. This suggests that there should be another
energy scale in the problem which was not considered. This scale, we
argue in this paper, is simply $T_K$. In fact, notice that one can
determine $T_K$ from the low voltage data, as the amplitude of the
tunneling conductance (which in Ref. \cite{Chang} was a fitting
parameter) is directly related to $T_K$ and the measured temperature
$T$:
\begin{equation}
G_0=\lim_{V\to 0}\ G=\nu\frac{e^2}{h}
\left[1-e^{-\frac{1}{2}\left(\frac{2\pi T}{T_K}\right)^2}\right]\ .
\end{equation}
In the case of the data in Ref. \cite{Chang}, $G_0\ll\nu e^2/h$, in
which case $T_K$ must be larger than the temperature scale, and
$G_0\approx \nu\frac{e^2}{h}{\frac{1}{2}\left(\frac{2\pi
T}{T_K}\right)^2}$. The same value of $T_K$ which is determined from
the low voltage data ($V\ll T)$ determines the energy scale for the
breakdown of the scaling form used in the fitting of the experimental
data. The breakdown voltage is $V\approx T_K$, and is taken into
account in Eq. (\ref{fullconductance}), which holds for the full range
of probed voltages.

The scaling form used in Ref. \cite{Chang} should break down when the
conductance becomes comparable to the natural conductance scale in the
problem, $\nu {\frac{e^2}{h}}$. This is explicit in
Eq. (\ref{fullconductance}). One should also interpret with care the
fact that the conductance saturates to the Hall value $\nu
{\frac{e^2}{h}}$. This is a point which should be taken into account
carefully, since for tunneling through a single impurity, as seen in
the previous section, the asymptotic value should be
$\frac{2\nu}{\nu+1}{\frac{e^2}{h}} >\nu {\frac{e^2}{h}}$.  What this
says is that for a small tunneling conductance, the multi-impurity
problem is very much like the single impurity one, but for large
tunneling conductance there must be a mechanism that lowers the
asymptotic value of the conductance to $\nu {\frac{e^2}{h}}$. This
mechanism is brought about by the multiple impurities.

\begin{figure}
\noindent
\epsfxsize=\linewidth
\epsfbox{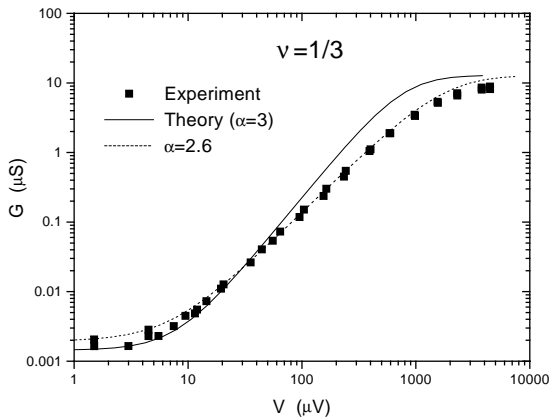}
\vspace{.2cm}
\caption{Comparison between experimental and theoretical conductances
for tunneling between an electron gas and a $\nu=1/3$ FQH state -
data set \#1. The values $T=25mK$ (in agreement with the measured
value) and $T_K=11K$ used in the theoretical ($g=2$,
$\alpha=2g-1=3$) plots are chosen to fit the low voltage end of the
curve. A better fit to the data is achieved using
Eq. (38) with $g=1.8$ (or $\alpha=2g-1=2.6$), $T=25mK$ and
$T_K=26.1K$. (Experimental data is courtesy of A. Chang.)}
\label{figdata1}
\end{figure}

\begin{figure}
\noindent
\epsfxsize=\linewidth
\epsfbox{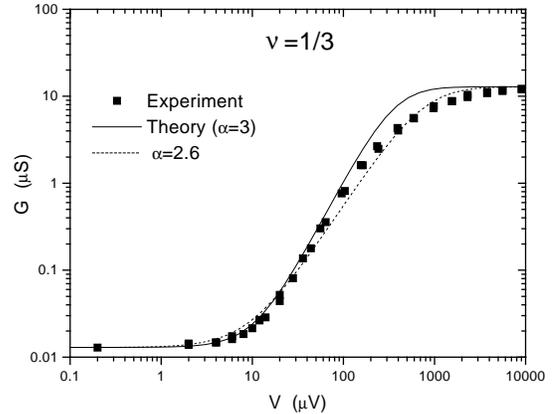}
\vspace{.2cm}
\caption{Comparison between experimental and theoretical conductances
for tunneling between an electron gas and a $\nu=1/3$ FQH state - data
set \#2. The set is best fit with a temperature $T=35mK$, higher than
the quoted $T=25mK$. The values of $T_K$ are $4.9K$ for the theoretical
($g=2$, $\alpha=2g-1=3$) curve, and $10.7K$ for the fit achieved using
Eq. (38) with $g=1.8$ (or $\alpha=2g-1=2.6$).
In this set there are more high voltage data points, which allows one
to see that the measured conductance saturates at $G=\frac{1}{3}
e^2/h$, in agreement with the theoretical prediction for
multi-impurity tunneling. This case is to be contrasted to the single
impurity value $G=\frac{1}{2} e^2/h$.  (Experimental data is courtesy
of A. Chang.)}
\label{figdata2}
\end{figure}

In Figures \ref{figdata1} and \ref{figdata2} we show, for $\nu=1/3$ in
two different samples, the comparison between the experimental data
for the conductance and the scaling form of
Eqs. (\ref{fullconductance}) and (\ref{fullotherg}), valid for the
full range of probed voltages.  We obtain the ratio $T_K/T$ from the
asymptotic low voltage conductance.  For the data set of
Fig.~\ref{figdata1} we use a temperature of $T=25mK$, the quoted
experimental value.  However, the data of Fig.~\ref{figdata2} appears
to be consistent with a higher temperature, close to $T=35mK$, rather
than $25 mK$. We would like to point out that our Eq.~\ref{smallVT},
which is a low voltage approximation of the full scaling form of the
conductance of Eq.~\ref{fullconductance}, is the same as the one used
to fit the data in ref.~\cite{Chang} with an exponent $\alpha=2g-1$.

The theoretical curve for $g=2$ (Eq.~(\ref{fullconductance}) (or
$\alpha=2g-1=3$) fits well the low-voltage data points, has a high
voltage crossover at about the right scale ($T_K$), and saturates to
the right high voltage asymptotic value ($\nu e^2/h$). However, for
both samples, it seems to overshoot the experimental data just above
the crossover energy scale. Instead, if we use Eq. (\ref{fullotherg})
with an effective exponent of $g=1.8$ ($\alpha=2g-1=2.6$), we get a
better fit to the data set over the full range of voltages. Notice,
however, that the value of $T_K$ for the two scaling forms changes by
about a factor of $2$ (for both samples).

The question on whether the exponent $g$ can be modified is an
important issue. In the case of tunneling from an electron gas into a
single chiral edge, the exponent cannot be modified since it is
determined by the dimension of the leading irrelevant operator and
should be $g=2$ ($\alpha=3$) for $\nu=1/3$.  An {\it effective}
exponent $\alpha \not= 3$ is indicative that there is an additional
crossover in the physics of these junctions.  A related issue is that
if we use $\alpha=3$, we find that the data is best fitted with values
of $T_K$ which are approximately half of those obtained for $\alpha
\approx 2.6$. This is a rather large change.  A possible explanation
is that subleading irrelevant tunneling operators are also coupled.
The main effects of such operators is to alter the corrections to
scaling. For example, the leading irrelevant operators cause an
effective $V$ dependence in the Kondo like temperature scale $T_K$.
In this case Eq. (\ref{fullconductance}), which assumes a $T_K$ that
does not scale, should be modified accordingly. Such a $V$ dependent
correction to $T_K$ could be a reason why other values of $g$ and
Eq. (\ref{fullotherg}) fit the experimental data better.  However,
other scenarios are also possible.  For instance, if there is clustering 
of tunneling centers at the atomic scale, the effective Hamiltonian will 
involve more that one impurity and more than one channel of the electron
gas. If that were the case, there would be additional
multi-impurity/multi-channel crossovers just above the Kondo scale but
will not affect neither the low voltage regime nor the high voltage
regime.

\subsection{Strong coupling}
Now let us discuss the case where the individual impurity couplings
are strong. In this case, we can still use the same recursion
Eq. (\ref{recursion})
with the strong coupling or high voltage ($V\gg T_K^{(n)}$) expression
for the current. Notice that what defines the strong coupling regime
is the ratio between the voltages and the energy scales set by the
impurities ($T_K^{(n)}$). Thus, for large enough voltages, tunneling
proceeds as if the coupling were strong, even if the nominal coupling
constants are finite. In this regime, for each individual impurity,
the conductance saturates at the value ${\tilde g}\frac{e^2}{h}=
\frac{2\nu}{\nu+1}\frac{e^2}{h}$, so that we have a recursion
\begin{eqnarray}
V_n-V_{n-1}&=&\nu^{-1}{\tilde g}\ (V_{R}-V_{n-1})\nonumber\\
&=&\frac{2}{\nu+1}\ (V_{R}-V_{n-1})\ ,
\end{eqnarray}
which has a solution
\begin{equation}
V_n=V_L+(V_R-V_L) \left[1-\left(\frac{\nu-1}{\nu+1}\right)^n\right]\ .
\end{equation}
Again, we have used that $V_0=V_L$ is the voltage at the incoming edge
branch, which is equilibrium with the other reservoir. Notice that the
voltage after scattering through the $n$-th impurity converges to
$V_\infty=V_R$, but not monotonically, oscillating around the
asymptotic value (because the argument being raised to the power $n$
is negative, since we have $\nu<1$). For example, in the case of
$\nu=1/3$, the sequence of voltages $(V_n-V_L)/(V_R-V_L)$ is
$\{0,3/2,3/4,9/8,15/16,\dots\}$, where we recognize $V_1$ as the
strong coupling result from the single impurity problem.

The conductance can be obtained from the current through the device,
$I=\nu (e^2/h)\ (V_N-V_0)$, which yields
\begin{equation}
G_N=\nu
\left[1-\left(\frac{\nu-1}{\nu+1}\right)^N\right]\ \frac{e^2}{h}\ .
\end{equation}
We thus have a sequence of quantized conductances converging to
the Hall value $\nu e^2/h$. For example, for $\nu=1/3$ the sequence is
\begin{equation}
G_N= \left\{
\frac{1}{2},\frac{1}{4},\frac{3}{8},\frac{5}{16},\dots,\frac{1}{3}
\right\}
\frac{e^2}{h}\ ,
\end{equation}
and for $\nu=1/5$
\begin{equation}
G_N= \left\{
\frac{1}{3},\frac{1}{9},\frac{7}{27},\frac{13}{81},\dots,\frac{1}{5}
\right\}
\frac{e^2}{h}\ .
\end{equation}
For $\nu=1$ we have the uninteresting and expected result that all
$G_N=e^2/h$, for all $N$.

In the above, we have assumed that the contact with the left reservoir
is made through many impurities, so that $V_0=V_L$. One can easily
treat in a very similar manner the case of $N_L$ contacts with the
left reservoirs, and $N_R$ contacts with the right one (as we did the
case of one contact for each in section \ref{sec:oneimp}). This
general case will display a conductance $G_{N_L,N_R}$ which will
depend on both $N_L$ and $N_R$:
\begin{equation}
G_{N_L,N_R}=
\frac{\left[1-\left(\frac{\nu-1}{\nu+1}\right)^{N_L}\right]
\left[1-\left(\frac{\nu-1}{\nu+1}\right)^{N_R}\right]}
{\left[1-\left(\frac{\nu-1}{\nu+1}\right)^{N_L+N_R}\right]}
\ \nu\frac{e^2}{h}\ .
\end{equation}
One can check that this formula, in particular, reproduces the cases
in Fig. \ref{fig1}: $G_{\infty,\infty}=\nu e^2/h$,
$G_{\infty,1}=\frac{2\nu}{\nu+1} e^2/h$, and $G_{1,1}= e^2/h$.

\subsection{Large number of impurities and equilibration}

One common conclusion to the case of weakly or strongly coupled
individual impurities is that, for a large number of them, the outgoing
edge equilibrates with the reservoir. How the series $V_n$ reaches the
asymptotic value $V_R$ may vary, but if the series converge, it must
converge to $V_R$ for a large number of impurities $N$. Indeed, this
result is a statement that the attractor for the recursion relation
Eq. (\ref{recursion}) is a single fixed point. The recursion
Eq. (\ref{recursion}) implies that, if the voltages $V_n$ converge,
then $V_\infty=V_R$, independent of the impurity couplings.

This is the mechanism for equilibration between the edge and the
reservoir. It had been noticed by Kane and Fisher that impurity
tunneling was key in this equilibration, in the case where one can
assume that the effect of tunneling is a small leaking conductance
which would equilibrate the edge and reservoir past an equilibration
length dependent on the leak conductance. What we show here is that
this equilibration mechanism is sturdier and more complex,
equilibrating edge and reservoir for very general coupling
distributions. The most striking example is that of strong coupling,
where the asymptotic voltage is reached non-monotonically, which is
important to understand how the result of a single impurity (hot
electrons) is possible, and can be reconciled with the idea of
tunneling causing equilibration.

\section{Conclusions}
\label{sec:conc}

In this paper we have investigated the problem of tunneling from an
electron gas (reservoir) to a FQH state for different types of
contacts.  We showed that different {\it universal} values can be
obtained for the two-terminal conductance.  At large voltages, or
strong coupling, the conductance of a point-like tunneling junction
between an electron gas reservoir and a Laughlin FQH state at filling
fraction $\nu$ was shown to {\it saturate} to a universal value
$G={\frac{2\nu}{\nu+1}}{\frac{e^2}{h}}$.  We used this result to show
that devices with different types of contacts between the reservoir
and the FQH state lead to distinct universal values of saturation
conductance which are rational multiples of $e^2/h$.  In particular,
the fraction $e^2/h$ was obtained for the case of electron tunneling
in and out of a FQH liquid by two point contacts.  We demonstrated
that the problem of tunneling between an electron gas and a fractional
quantum Hall state through an impurity is exactly equivalent to the
problem of tunneling between a chiral Fermi liquid and a chiral
Luttinger liquid.  The interesting case of tunneling to a $\nu=1/3$
FQH state was investigated in detailed and shown to be equivalent to
the problem of tunneling {\it between} two $g=1/2$ chiral Luttinger
liquids. This system provides an experimental realization of this
important exactly solvable case.  The results of the single impurity
problem were used to consider the case of many tunneling centers
coupled independently to an electron reservoir. This problem is
relevant to recent experiments by A.~Chang {\it et.~al.} Using the
exact solution of the single impurity problem, we derived an explicit
universal expression for the voltage and temperature dependent
conductance for a problem of many independent impurities. Here we made
the key assumption that the channels of the electron gas which couple
to each individual impurity are always in equlibrium or, what is the
same, that there is no phase coherence between channels. This
assumption must be accurate for widely separated tunneling impurities
of an atomically smooth junction.  We showed that the voltage and
temperature dependent conductance exhibits a crossover reminiscent of
a Kondo effect with a Kondo scale $T_K$ determined by the tunneling
matrix matrix element.  This universal curve was shown to fit the
experimental data over the full range of probed voltages.  It was also
observed that this universal curve oveshoots the experimental data on
a voltage scale above $T_K$ but below saturation where an effective
exponent was found to give a better fit to the data.  We interpreted
this effective exponent as indicating that either a subleading
irrelevant operator had a significant amplitude (which would be the
case if the edge structure is not sharp) or that the physical samples
had some degree of clustering of tunneling centers, leading to
multi-channel/multi-impurity physics. Also, in this work we have
assumed that the tunneling matrix element $\Gamma$ is independent of
the applied voltage. Clearly, as $V$ increases $\Gamma$ should change
\cite{ChristenandButtiker}. However, these changes amount to an
analytic redefinition of the coupling constant and are
non-universal. Such effects lead to a redefinition of $T_K$, and do
not change either the exponent or the saturation conductance.

\begin{center}
{\bf ACKNOWLEDGEMENTS}
\end{center}

We are particularly grateful to Albert Chang for several
enlightening discussions, as well as for making his data available to us. 
We would also like to thank Matthew Grayson and Nancy Sandler for 
many useful comments. This
work was supported in part by the National Science Foundation through
the grants NSF DMR94-24511 at the University of Illinois at
Urbana-Champaign and NSF DMR-89-20538 at the Materials Research
Laboratory of the University of Illinois at Urbana-Champaign.

\appendix
\section{General map from a high dimensional electron gas to a 1D
chiral Fermi liquid for tunneling through a single impurity}
\label{appendiximp}

Here we consider in detail the general problem of coupling an
impurity to an electron gas subject to generic boundary conditions. We
show that, regardless of details of the boundary conditions, only one
quantum channel couples to the impurity, and the electron gas can be
regarded as a 1D chiral Fermi liquid as what concerns the impurity
coupling.

The sufficient assumption that we make is that the electron gas 
in the bulk is
isotropic, so that the energy  $\epsilon({\vec k})$ 
of the bulk eigenstates
depends only on $|{\vec k}|$. In this case we can expand the electron
operator as
\begin{equation}
\psi(\vec x)=\int_0^\infty dk \sum_{\bf \lambda}
\ \phi_{k,{\bf \lambda}}(\vec x)
\ c_{\bf \lambda}(k)\ ,
\end{equation}
where $\{\phi_{k,{\bf \lambda}}(\vec x)\}$ is a complete set on
one-particle eigenstates which satisfy the correct boundary conditions.
Here $\{{\bf \lambda}\}$ is a set of quantum numbers which label {\it
degenerate} states with the same wavenumber $k$ and $\epsilon(k)$. 
The electron operator obey the anti-commutation relation
\begin{equation}
\{c^\dagger_{\bf \lambda}(k),c_{\bf \lambda'}(k')\}=
\delta_{{\bf \lambda},{\bf \lambda'}}\
\delta(k-k')\ .
\end{equation}

At the impurity location, which we take to be ${\vec x}=0$, we have
\begin{equation}
\psi(0)=\int_0^\infty dk \sum_{\bf \lambda}
\ \phi_{k,{\bf \lambda}}(0)
\ c_{\bf \lambda}(k)\ ,
\end{equation}
and because the states with the same $k$ are degenerate, we can perform
an orthonormal transformation so as go to a new basis where
\begin{equation}
{\tilde c}_{\bf \alpha}(k)={\frac{1}{{\cal N}_k }}
\sum_{\bf \lambda}
\ \phi_{k,{\bf \lambda}}(0)
\ c_{\bf \lambda}(k)
\end{equation}
is a basis vector. The normalization factor ${\cal N}_k$
appears because the vector has not necessarily norm 1 ($\psi(0)$ has
weight at different $k$).
This can always be done via a Gram-Schmidt
orthogonalization process. Notice that the fact that there is a single
impurity is used here: one impurity picks only one direction in each 
subspace labelled by
$k$. Hence, we can always use this one direction as the first basis 
vector in the Gram-Schmidt process. 

We can thus write
\begin{equation}
\psi(0)=\int_0^\infty dk\ {\cal N}_k
\ {\tilde c}_{\bf \alpha}(k)\ ,
\end{equation}
which displays clearly that the impurity only couples to a single
channel (${\bf \alpha}$). The fermion operators in this channel
satisfy $\{{\tilde c}^\dagger_{\bf \alpha}(k),{\tilde c}_{\bf
\alpha}(k')\}= \ \delta(k-k')$.

The coupled (${\bf \alpha}$) channel can be described in terms of
non-chiral fermions in a semi-infinite line, where left and right
moving particles correspond to incoming and outgoing particles with
respect to the impurity. This half-line can be unfolded, so we are
left with one chiral fermion on an infinite line. Thus, from the
perspective of the impurity, the electron gas can be regarded as a
chiral Fermi liquid. One should notice that $g=1$ is completely fixed
in this problem, because the chiral fermions are derived from a higher
dimensional system, where the Fermi liquid picture holds.

Below we give particular examples which are applications of our
general result.

\subsection{Spherically symmetric system}

This case is a simple application of the general result. Here we
follow closely the derivation for the case of the impurity at the bulk
by Affleck and Ludwig \cite{Affleck&Ludwig}.

In this case, one writes the electron operator in terms of plane waves
as
\begin{equation}
\psi(\vec x)=\int\frac{d^3k}{(2\pi)^{3/2}}\ e^{i\vec k\cdot \vec x}\
c_{\vec k}\ ,
\end{equation}
where the operators $c_{\vec k}$ satisfy the anti-commutation
relations $\{c^\dagger_{\vec k},c_{\vec k'}\}=
\delta^3({\vec k}-{\vec k'})$. One then notices that at the impurity
location $\vec x=0$,
\begin{equation}
\psi(0)=
\int\frac{d^3k}{(2\pi)^{3/2}}\ c_{\vec k}
\end{equation}
depends only on the spherically symmetric component ($L=0$) of the
operator $c_{\vec k}$, namely
\begin{equation}
c_{L=0}(k)=\frac{k}{\sqrt{4\pi}}\int d\hat\Omega\ c_{\hat\Omega k}\ ,
\end{equation}
which satisfy $\{c^\dagger_{L=0}(k),c_{L=0}(k')\}=\delta(k-k')$. Thus,
one has only to consider the 1-D fermions running along the radial
coordinates for the $L=0$ mode for the purposes of coupling to the
impurity at $\vec x=0$.

\subsection{Impurity at a planar boundary of a 3D electron gas}

we now consider in detail the case when the point-like contact or
impurity is not in the bulk, but at the planar boundary between a free
electron gas and a potential barrier that confines the gas. Here we
have to take into account that the eigenstates are modified by the
presence of the boundary.

Let the boundary be at the $z=0$ plane. Let the barrier high be $U>0$
for $z<0$, and 0 for $z>0$. The effect of the barrier on the
wavefunctions can be absorbed completely in the phase shift that the
electron picks up after reflecting of the boundary (the barrier is
confining, so the Fermi level lies under the barrier height, and thus
the barrier is completely reflecting). The wavefunctions for $z>0$ can
then be written as
\begin{equation}
\phi^{(+)}_{{\vec k}_\perp,k_z}(\vec x)=e^{i{\vec k}_\perp\cdot x_\perp}\
\frac{e^{-i\frac{1}{2}\phi(k_z)}\ e^{i k_z z}+
e^{i\frac{1}{2}\phi(k_z)}\ e^{-i k_z z}}{\sqrt{2}}\ ,
\end{equation}
where $e^{i\phi(k_z)}$ is the phase shift factor due to the reflection
at the boundary of a wave with momentum $k_z$ (notice that
$\phi(k_z)=-\phi(-k_z)$). It is an elementary exercise to show that
for a potential barrier of of height $V$ the phase shift is given by
$\cos(\phi(k_z)/2)=|{k_z}/{k_0}|$, where $k_0=\sqrt{2mU}/\hbar$.

The operator that creates the state
$\phi^{(+)}_{{\vec k}_\perp,k_z}(\vec x)$
can be written in terms of plane wave creation
operators $c^\dagger_{\vec k}$ as
\begin{equation}
{\gamma^{(+)}}^\dagger_{{\vec k}_\perp,k_z}=
\frac{e^{-i\frac{1}{2}\phi(k_z)}\ c^\dagger_{{\vec k}_\perp,k_z}+
e^{i\frac{1}{2}\phi(k_z)}\ c^\dagger_{{\vec k}_\perp,-k_z}}{\sqrt{2}}\ .
\end{equation}
Although the general result we have shown has a simple proof, the
aplication to a particular case involves explicitly finding the right
basis. In this particular problem at hand, this can be greatly
simplified by exploring symmetries and enlarging the Hilbert space.

The operators ${\gamma^{(+)}}^\dagger_{{\vec k}_\perp,k_z}$ generate
only half the Hilbert space for free fermions without the
boundary. The other half is generated from wavefunctions of different
symmetry
\begin{equation}
\phi^{(-)}_{{\vec k}_\perp,k_z}(\vec x)=e^{i{\vec k}_\perp\cdot x_\perp}\
\frac{e^{-i\frac{1}{2}\phi(k_z)}\ e^{i k_z z}-
e^{i\frac{1}{2}\phi(k_z)}\ e^{-i k_z z}}{\sqrt{2}}\ ,
\end{equation}
by operators
\begin{equation}
{\gamma^{(-)}}^\dagger_{{\vec k}_\perp,k_z}=
\frac{e^{-i\frac{1}{2}\phi(k_z)}\ c^\dagger_{{\vec k}_\perp,k_z}-
e^{i\frac{1}{2}\phi(k_z)}\ c^\dagger_{{\vec k}_\perp,-k_z}}{\sqrt{2}}\ .
\end{equation}
We can redefine fermion operators ${\tilde c}_{\vec k}={\rm
sgn}(k_z)\ e^{i\frac{1}{2}\phi(k_z)}\ c_{\vec k}$ so as absorb the
phases into the free fermions.  The Hamiltonian for the system is just
that for free fermions $c_{\vec k}$:
\begin{equation}
H=\int\frac{d^3k}{(2\pi)^{3/2}}\ \frac{\hbar^2 k^2}{2 m}
\ c^\dagger_{\vec k} c_{\vec k}=
\int\frac{d^3k}{(2\pi)^{3/2}}\ \frac{\hbar^2 k^2}{2 m}
\ {\tilde c}^\dagger_{\vec k} {\tilde c}_{\vec k}
\end{equation}

We can write the electron operator for $z\ge 0$ as
\begin{equation}
\psi(\vec x)=\int\frac{d^2k_\perp}{(2\pi)}
\int_0^\infty \frac{dk_z}{(2\pi)^{1/2}}\
\phi^{(+)}_{{\vec k}_\perp,k_z}(\vec x)\
{\gamma^{(+)}}_{{\vec k}_\perp,k_z}\ .
\end{equation}
At the impurity location $\vec x=0$,
\begin{eqnarray}
\psi(0)&=&
\int\frac{d^2k_\perp}{(2\pi)}
\int_0^\infty \frac{dk_z}{(2\pi)^{1/2}}\
\phi^{(+)}_{{\vec k}_\perp,k_z}(0)\
{\gamma^{(+)}}_{{\vec k}_\perp,k_z}\nonumber\\
&=&
\int\frac{d^3k}{(2\pi)^{3/2}}\ \cos\frac{\phi(k_z)}{2}\  {\rm sgn}(k_z)
\ {\tilde c}_{\vec k}\nonumber\\
&=&
\int\frac{d^3k}{(2\pi)^{3/2}}
\ {\tilde c}_{\vec k}\ \frac{k_z}{k_0}\ ,
\nonumber
\end{eqnarray}
which depends only on the ($L=1,M=0$) mode of the
operator ${\tilde c}_{\vec k}$, namely
\begin{equation}
{\tilde c}_{L=1,M=0}(k)=
k\sqrt{\frac{3}{4\pi}}
\int d\hat\Omega\ {\tilde c}_{\hat\Omega k}\ \cos\theta\ ,
\end{equation}
which satisfy the commutation relations
\begin{equation}
\{{\tilde c}^\dagger_{L=1,M=0}(k),{\tilde c}_{L=1,M=0}(k')\}=\delta(k-k')\ .
\end{equation}
Again, one has only to consider the 1-D fermions running along the
radial coordinates for this angular momentum mode.






%
%
%
%
%
%

\end{multicols}

\end{document}